# Effect of electron doping on the magnetic correlations in the bilayered brownmillerite compound $Ca_{2.5-x}La_xSr_{0.5}GaMn_2O_8$: a neutron diffraction study

A. K. Bera and S. M. Yusuf[*]

Solid State Physics Division, Bhabha Atomic Research Centre, Mumbai 400 085, India

I. Mirebeau

Laboratoire Léon Brillouin, CEA-CNRS, CEA/Saclay, 91191 Gif sur Yvette, France

**Abstract**

The effect of electron doping on the magnetic properties of the Brownmillerite-type bilayered compounds has been investigated by neutron powder diffraction in La substituted $Ca_{2.5-x}La_xSr_{0.5}GaMn_2O_8$ compounds ($x = 0.05$ and $0.1$), in comparison with the undoped compound ($x = 0$). In all compounds, a long-range three dimensional collinear antiferromagnetic (AFM) structure is found below the Néel temperature $T_N$ of respective compound, whereas, well above $T_N$, three dimensional short-range magnetic ordering is observed. In the intermediate temperature range just above $T_N$, a strong effect of electron doping (La substitution) on the magnetic correlations has been observed. Here, a short-range AFM correlation with possible dimensionality of three has been found for substituted compounds ($x = 0.05$ and $0.1$) as compared to the reported two dimensional long-range AFM ordering in the parent compound. With increasing the electron doping, a decrease in $T_N$ is also observed. The short-range magnetic correlations set in over a large temperature range above $T_N$. A magnetic phase diagram in the $x$-$T$ plane is proposed from these results.

1. Introduction

Magnetic ordering in a reduced dimension ($D < 3$), and its crossover to a higher dimension are fundamental issues in condensed matter physics [1, 2]. They can be mostly due to strongly anisotropic exchange interactions, or to the strongly frustrating topology of systems. The macroscopic degeneracy of the magnetic ground state in low dimensional magnetic systems makes such systems very sensitive to an additional exchange interaction, leading to various unconventional magnetic properties. Similarly, the dimensionality of the magnetic ordering and its crossover to a higher dimension are very sensitive to external perturbations, such as temperature, magnetic field, and pressure, as well as to internal perturbations like substitutions, inhomogeneities, *etc*. A dimensionality crossover of the magnetic ordering from two dimensional (2D) to three dimensional (3D) with lowering of temperature has been observed in various kinds of 2D systems, such as bilayered $Ca_{2.5}Sr_{0.5}GaMn_2O_8$ compound [3], parent systems of high-$T_C$ superconducting cuprates ($La_2CuO_4$ and $YBa_2Cu_2O_6$) [4, 5], layered borocarbides $RB_2C$ ($R$=Dy, Ho, and Er) [6], layered manganites $La_{3-x}Sr_xMn_2O_7$ [7], and high-temperature superconducting oxides ($ErBa_2Cu_4O_8$, $DyBa_2Cu_4O_8$, and $Pb_2Sr_2PrCu_2O_8$) involving rare earth magnetism [8-11]. A pressure induced crossover from 2D to 3D magnetic correlations has been observed in $La_{1.4}Sr_{1.6}Mn_2O_7$ manganite [12]. A field induced 3D magnetic ordering has been reported in quasi 1D spin-gap systems [13-15]. A strengthening of the 2D magnetic ordering has been observed with Ba substitution in layered oxysulfide $(Sr,Ba)_2Cu_2CoO_2S_2$ due to an increase of the separation length between the magnetic $CoO_2$ layers, which arises because of the difference between ionic radii of smaller $Sr^{2+}$ (1.26 Å) and larger $Ba^{2+}$ (1.42 Å) [16]. Similarly, investigation of an electron doping effect on the magnetic correlations/dimensionality is also an important and interesting topic in low dimensional magnetism. An electron doping can be performed by a suitable chemical substitution having equal ionic radius with higher oxidation state. Due to the equal ionic radii of the ions, substitution does not have any effect on crystal structure.

In this paper, we study the effect of electron doping on the magnetic correlations in the bilayered compounds $Ca_{2.5-x}La_xSr_{0.5}GaMn_2O_8$ ($x$ = 0.05 and 0.1). Here, the electron doping was done by substitution of $La^{3+}$ (1.16 Å), with almost equal ionic radius, at the $Ca^{2+}$ (1.12 Å) site. The substitution causes a change in the oxidation state of Mn ions from +4 to +3 and as a result, doping of electrons occurs in the system. These $Ca_{2.5-x}La_xSr_{0.5}GaMn_2O_8$ compounds belong to the family of bilayered Brownmillerite compounds of the type $A_3B'B_2O_8$ [$A$ = La, Ca, Sr, Y, and ($B'$, $B$) = alkali or transition metal ions]. They are of particular interest due to their unusual low dimensional magnetic properties as well as their possible applications in spintronics [3, 17-22]. These compounds crystallize in the orthorhombic crystal structure (space group $Pcm2_1$) with an alternating stacking of $MnO_6$ octahedral bilayer and a single layer of $GaO_4$ tetrahedra along the crystallographic $b$ direction. The cations $A$ occupy the positions between these layers in a disordered manner [17, 18]. The magnetic properties of the parent compound $Ca_{2.5}Sr_{0.5}GaMn_2O_8$ show a 3D short-range (SR) antiferromagnetic (AFM) ordering over 180-170 K, a 2D long-range (LR) AFM ordering over 165-160 K, and then a 3D LR AFM ordering below ~ 160 K with an ordered magnetic moment of 3.09 (1) $\mu_B$ per Mn cation at 5 K (aligned along [010]) [3, 19]. The undoped compound $Ca_{2.5}Sr_{0.5}GaMn_2O_8$ is purely AFM, whereas ferromagnetic (FM) interactions, mediated by double exchange mechanism, start to occur with La-substitution. As a result, a ferromagnetic-like magnetic behaviour with enhanced magnetoresistance has been observed [17]. These properties make this type of naturally grown layered systems suitable for spintronics applications. The observed changes in the bulk magnetic and electronic properties with La substitution are due to the electron doping i.e., the increase in the ratio between $Mn^{3+}$ and $Mn^{4+}$. Our previous crystal structural study based on room temperature X-ray diffraction suggested that there was no observable change in the average crystal structure with La substitution [17]. However, an enhancement in the local Jahn-Teller distortion may be expected due to the increase of the concentration of $Mn^{3+}$ ion with La-substitution. In the present work, we have used the neutron diffraction technique to investigate microscopically the dimensionality of the magnetic ordering and the spin–spin correlation length in the electron doped compounds $Ca_{2.5-x}La_xSr_{0.5}GaMn_2O_8$ ($x$ = 0.05 and 0.1). Most interestingly, by studying the magnetic correlations in these compounds just above the Néel temperature for the onset of 3D long-range order, our study demonstrates a strong effect of electron doping on the magnetic correlations, where a two dimensional long-range AFM ordering for $x$ = 0 compound changes to a short-range AFM ordering with possible dimensionality of three in the doped compounds ($x$ = 0.05 and 0.1).

2. **Experimental details**

The polycrystalline samples of $Ca_{2.5-x}La_xSr_{0.5}GaMn_2O_8$ ($x$ = 0.05 and 0.1) were prepared by the solid state reaction method as described in our previous report [17]. Neutron powder diffraction measurements were carried out over the temperature range of 4.5-275 K using the G6.1 powder neutron diffractometer at Laboratoire Léon Brillouin, Saclay with a wavelength of $\lambda$ = 4.751 Å, covering a scattering vector length [$Q$= (4π/sinθ)] of ~ 0.120–2.40 Å$^{-1}$. For these measurements, the samples were placed in aluminium cans and a commercial liquid helium cryostat was used to vary the temperature. The diffraction data were analyzed by the Rietveld method using the FULLPROF program [23, 24]. The representation theory analysis was performed using the BASIREPS software for the determination of the magnetic structure [23, 24].

## 3. Results and discussion

Figures 1(a) & 1(b) show the observed and Rietveld refined neutron powder diffraction patterns for the compounds $Ca_{2.5-x}La_xSr_{0.5}GaMn_2O_8$ with $x$ = 0.05 and 0.1, respectively, at 275 K (upper panel) and 4.5 K (lower panel). Rietveld analysis of the diffraction patterns confirms the single phase state of both samples with an orthorhombic crystal structure (Space group $Pcm2_1$). The refined lattice parameters are given in table 1. The values of the lattice parameters are in good agreement with the previously reported values, obtained from an independent X-ray diffraction study [17]. The diffraction patterns at 275 K could be fitted with only nuclear intensities confirming the paramagnetic nature of the samples at this temperature. The fractional atomic positions were taken from our earlier X-ray diffraction study [17], and kept fixed during refinement. This method was adopted since the available $Q$-range of the present diffraction patterns is limited ($Q$ ~ 0.15 – 2.5 Å$^{-1}$; scattering angular range ($2\theta$) = 5-120º) and does not allow one to derive reliable positional parameters and temperature factors. The temperature factors were therefore kept fixed to the values derived for the iso-structural compound $Ca_{2.5}Sr_{0.5}(Ga,Co)_{1+x}Mn_{2-x}O_8$ [Ref 21] and this should not affect the magnetic model used in the present study (described later). During the refinement, only the lattice constants and profile parameters were varied. As shown in figure1, the appearance of additional magnetic Bragg peaks at $Q$ ~ 1.20, 1.32 and 1.63 Å$^{-1}$ along with an increase in the intensity of the (100), (110), and (120) [$Q$ ~ 1.17, 1.29, and 1.61 Å$^{-1}$] fundamental (nuclear) Bragg peaks confirms a 3D AFM ordering at 4.5 K. For both compounds, the same type of AFM structure is found. All magnetic peaks can be indexed with a propagation vector $k$ = (0 0 0); this indicates an AFM structure having the same unit cell as the chemical one. The magnetic structure has been analyzed using the standard irreducible representational theory as described by Bertaut [25, 26]. For the propagation vector $k = (0\ 0\ 0)$, the irreducible representations of the propagation vector group $G_k$ are given in table 2. In the space group $Pcm2_1$ with the propagation vector $k = (0\ 0\ 0)$, there are four possible irreducible representations. The magnetic reducible representation $\Gamma$ for $4c$ site (Mn-site) can be decomposed as a direct sum of irreducible representations as

$\Gamma_{mag} = 3\ \Gamma_1 + 3\ \Gamma_2 + 3\ \Gamma_3 + 3\ \Gamma_4$

The basis vectors of the Mn position $4c$ $(x, y, z)$ for the representations, calculated using the projection operator technique, implemented in BASIREPS [24], are given in table 3. Among all four representations, the refinement of the magnetic structure with the representation $\Gamma_1$ gives the best fit to the observed diffraction patterns at 4.5 K, as shown in the lower panels of figures 1(a) and 1(b) for the compounds $x = 0.05$ and 0.1, respectively. The corresponding magnetic structure is shown in figure 2(a). Here, Mn spins are coupled antiferromagnetically in a given $ac$ plane (single layer of $MnO_6$). A ferromagnetic coupling of such AFM $ac$ planes is found along the $b$ axis within a bilayer. Moreover, bilayers in adjacent unit cells, separated by non-magnetic $GaO_4$ layer, are coupled ferromagnetically along the $b$ axis. The Mn magnetic moments align along the [010] crystallographic direction. The absence of a magnetic contribution on the $(0k0)$ nuclear (fundamental) Bragg peaks, such as (010) and (020) which are temperature independent [figures 2(b) and 2(c)], confirms the alignment of the Mn magnetic moments along the [010] axis. The spin arrangement is consistent with earlier reports for the parent compound $Ca_{2.5}Sr_{0.5}GaMn_2O_8$ [3, 19]. A theoretical calculation using the spin-polarized generalized gradient approximation method and ultrasoft pseudopotentials for the parent compound also showed that the observed AFM spin configuration was the most stable one for this system [27]. At 4.5 K, the site averaged ordered magnetic moment is found to be 3.02(1) and 3.01(1) $\mu_B$/Mn ion for the compounds with $x = 0.05$ and 0.1, respectively. The values of the ordered moments are consistent with the moment value reported for the parent compound [3.09(1) $\mu_B$/Mn at 5 K]. The *R* factors for the refined neutron diffraction patterns at 4.5 K are: $R_{Nuclear}$ = 4.11% and $R_{Magnetic}$ = 3.20% for the sample *x* = 0.05 and $R_{Nuclear}$ = 6.26%, $R_{Magnetic}$ = 4.51% for the sample *x* = 0.1, respectively (table 1).

Figure 3 shows the evolution of the magnetic moment with temperature for *x* = 0.05 (solid circle) and 0.1 (open square) samples. The $T_N$ are found to be ~150 K and ~ 142.5 K for the samples with *x* = 0.05 and 0.1, respectively. The value of $T_N$ for the parent compound was reported to be ~ 160 K [3, 18]. Therefore, a decrease of $T_N$ is evident with the La substitution. Here, we would like to mention that our previous study using magnetization and magnetoresistance showed the appearance of FM interactions due to double exchange (DE) mechanism with electron doping or La substitution [17]. A competition between the leading superexchange (SE) AFM and the electron doping induced double exchange (DE) FM interactions may cause a decrease in the average SE AFM exchange interaction, resulting to the observed decrease of the $T_N$ value.

To understand the nature of the 3D long-range AFM ordering in more details, we have plotted the temperature dependence of the half width at half maximum (HWHM) of the two most intense magnetic Bragg peaks (100) and (001). In figure 4, the averaged values of HWHM of these two peaks are shown as a function of temperature for the *x* = 0.05 (open circle) and 0.1 (open square) compounds. The widths of the magnetic peaks remain unchanged only below ~ 135 and ~ 125 K for the samples with *x* = 0.05 and 0.1, respectively, when they are limited by the instrumental resolution. However, as discussed earlier, the 3D ordering temperatures ($T_N$) are found to be ~150 K and ~ 142.5 K for *x* = 0.05 and 0.1 samples, respectively [figure 3]. Therefore, the true 3D long-range AFM ordering occurs only at $T \leq$ 135 K (for *x* = 0.05) and 125 K (for *x* = 0.1), respectively when the width of the magnetic Bragg peaks are limited by the instrumental resolution. A broadening of the peak widths over ~ 135-150 K for the *x* = 0.05 and ~ 125-142.5 K for the *x* = 0.1 samples indicates a decrease of the magnetic correlation length with increasing temperature up to $T_N$. This type of ordering could be called a restricted long-range (RLR) 3D ordering. It should be noted that this RLR type ordering over the temperature range of 140-160 K was also observed for the parent compound, in which true 3D LR magnetic ordering occurs only below 140 K [3]. Therefore, the La-substitution in the layered system $Ca_{2.5}Sr_{0.5}GaMn_2O_8$ results in a decrease in the 3D AFM ordering temperature.

Figures 5(i) and 5(ii) show the purely magnetic diffraction patterns above $T_N$ (at 200, 185, 170, 155, and 152.5 K) for the $x = 0.05$ and (at 200, 185, 170, 155, and 145 K) 0.1 samples, respectively. These magnetic patterns are obtained after subtraction of a diffraction pattern measured at high temperature (210 K), in the paramagnetic state. For the $x = 0.05$ sample, with decreasing temperature below 195 K, a broad peak starts to appear at a scattering vector of $Q \sim 1.17$ Å$^{-1}$. In the temperature range 190-155 K, the peak intensity increases with decreasing temperature. When further lowering the temperature, the broad peak slowly transforms into two asymmetric Bragg peaks centred at $Q \sim 1.15$ and 1.18 Å$^{-1}$ at 152.5 K. As described in the previous paragraph, below $T_N=150$ K two symmetric Bragg peaks (termed as RLR), corresponding to the 3D magnetic Bragg peaks (100) and (001), appear at the same $Q$ positions $\sim 1.15$ and 1.18 Å$^{-1}$, respectively. For the $x = 0.1$ sample also, a broad peak (centred at $Q \sim 1.18$ Å$^{-1}$) appears below 195 K. With decreasing temperature down to $T_N$ (142.5 K), the peak profiles remain same with a gradual increase in the intensity. Below $T_N$, two symmetric Bragg peaks (termed earlier as RLR), corresponding to the 3D magnetic Bragg peaks (100) and (001), appear at $Q \sim 1.17$ and 1.2 Å$^{-1}$, respectively [figures 1 and 4]. The peak positions ($\sim 1.17$ and 1.18 Å$^{-1}$ for $x = 0.05$ and 0.1, respectively) suggest similar AFM spin periodicities in both samples. But the peak profiles at 152.5 K ($T/T_N \sim 1.02$) for $x = 0.05$ and at 145 K ($T/T_N \sim 1.02$) for $x = 0.1$ are quite different, suggesting different types of spin-spin correlations in these samples.

We first study the peak profile of the $x = 0.05$ sample at 152.5 K. An asymmetric saw-tooth type peak profile occurs when scattering takes place from a randomly stacked 2D ordered system [3, 8-10, 28-30]. On the other hand, an almost symmetric type peak profile occurs when the scattering takes place from a 3D ordered system. The scattering intensity of a 2D Bragg reflection ($hl$) can be expressed by the Warren function[31] as

$$I_{hl}(2\theta) = C \left[ \frac{\xi_{2D}}{(\lambda\sqrt{\pi})} \right]^{1/2} j_{hl} |F_{hl}|^2 \frac{(1+\cos^2 2\theta)}{2(\sin^{1/2}\theta)} F(a) \qquad (1)$$

where, $C$ is a scale factor, $\xi_{2D}$ is the 2D spin-spin correlation length within the 2D layer, $\lambda$ is the wavelength of the incident neutrons, $j_{hl}$ is the multiplicity of the 2D reflection ($hl$) with 2D magnetic structure factor $F_{hl}$, and $2\theta$ is the scattering angle. The function $F(a)$ is given by $F(a) = \int_0^\infty \exp\left[-(x^2-a)^2\right] dx$ where, $a = (2\xi_{2D}\sqrt{\pi}/\lambda)(\sin\theta - \sin\theta_{2DB})$ and $\theta_{2DB}$ is the Bragg angle for the 2D ($hl$) reflection. On the other hand, the scattering intensity due to 3D SR spin-spin correlations can be expressed by a Lorentzian function as

$$I_{hkl}(Q) = A / \left[(Q-Q_C)^2 + \kappa^2\right] \qquad (2)$$

Here $A$ is the proportionality constant, $\kappa$ is the inverse of the 3D magnetic correlation length $\xi_{3D}$ and $Q_c$ is the length [$Q = (4\pi \sin\theta)/\lambda$] of the scattering vector at which the peak of $I(Q)$ occurs. For the $x = 0.05$ sample, the observed asymmetric type peak profile at 152.5 K could not be fitted with Lorentzian function alone. The peak profile can, however, be reproduced by considering a combined function of both Warren and Lorentzian functions. We calculated numerically the scattering intensity by considering both Warren (Eq. 1) and Lorentzian (Eq. 2) functions [thick red curves in figures 5 (i) (e)]. The additional two magnetic peaks ~ $Q = 1.3$ Å$^{-1}$ have also been considered for calculation. A good agreement between calculated and observed peak profiles has been observed. However, one may argue that the observed asymmetric peak profile $Q \sim 1.15$ and 1.18 Å$^{-1}$ at 152.5 K (corresponding to the Warren function) may arise due to the contribution of other two magnetic peaks at $Q \sim 1.3$ Å$^{-1}$. To rule out this possibility, we have shown, in the inset of figure 5 (i)-(e), the fitting of four Lorentzian functions centred at the position of the Bragg peaks (100), (001), (110), and (011) as expected for a 3D magnetic ordering. This model does not represent the observed asymmetric profile. The above analysis suggests that for the $x = 0.05$ sample at 152.5 K a short-range magnetic ordering is present. The nature of short-range ordering could be a mixed 2D and 3D type. However, the observed asymmetric broadening (which is analyzed with a 2D-model) may arise from Jahn-Teller lattice distortions in Mn$^{3+}$O$_6$ octahedral sites of the doped compounds that create additional magnetic scattering (discussed later in detail). The values of the 2D correlation length ($\xi_{2D}$) and 3D correlation length ($\xi_{3D}$) were found to be 435(15) and 20(2) Å, respectively. For the $x = 0.05$ compound, at the higher temperatures ($T \geq 155$ K), the observed broad scattering profiles have been analyzed using a Lorentzian function. On the other hand, for the other compound ($x = 0.1$), the observed broad peak can be fitted with a Lorentzian peak shape over the whole temperature range 145-185 K. The temperature evaluation of the $\xi_{3D}$ for both samples is shown in figure 6.

Now, we compare the magnetic correlations above $T_N$ for the doped compounds with that of the parent undoped compound. For the parent system, a broad peak centred at $Q \sim 1.16$ Å$^{-1}$, corresponding to a pure 3D short-range spin-spin correlations, was reported at higher temperatures (180-170 K) and could be fitted with a single Lorentzian function. With decreasing temperature (over 165–160 K), the broad peak was found to transform gradually to two saw-tooth type asymmetric Bragg peaks, corresponding to a pure 2D long-range magnetic ordering, centred at $Q \sim$ 1.16 and 1.19 Å$^{-1}$. Then, at $T < 160$ K (over 155–150 K), the saw-tooth type asymmetric peaks gradually disappeared and symmetric Bragg peaks, corresponding to a pure 3D long-range magnetic ordering, appeared at the same scattering angles. With electron doping, no pure 2D magnetic ordering is found in the present study; rather a short-range magnetic ordering occurs above $T_N$ for the La substituted compounds. Here we would like to point out that a pure 3D nature of the short-range magnetic ordering is more evident for the higher electron doped $x = 0.1$ compound. The diffuse scattering observed for the present compounds is purely magnetic in nature as evident from the following points. (i) The positions of the diffuse and magnetic Bragg peaks coincide, and (ii) the diffuse scattering intensity increases with lowering the temperature down to $T_N$ then transforms gradually into the magnetic Bragg peaks. However, an effect of a local Jahn-Teller distortion on the observed magnetic diffuse scattering is possible as with the La substitution, a local Jahn-Teller lattice distortion may occur in $Mn^{3+}O_6$ octahedral sites due to the increase of the $Mn^{3+}$ ions. The Jahn-Teller lattice distortion may induce an additional neutron scattering due to magnetic disorder in $Mn^{3+}$ sites of the substituted compounds. This information can be obtained from Rietveld analysis of neutron diffraction patterns by measuring the Mn-O distances as a function of temperature. However, in the present work, the limited $Q$-range of the diffraction patterns does not permit us to estimate the Mn-O bond lengths accurately, hence, the Jahn-Teller distortions.

Now, we focus on the nature of magnetic ordering around and below $T_N$. The pure magnetic patterns (after subtraction of the nuclear background at 210 K) at 155, 150, 145, 140, 135, and 125 K for $x = 0.05$ and at 145, 140, 135, 130, 125, and 100 K for $x = 0.1$ compounds are shown in figures 7 (a) and 7 (b), respectively. A coexistence of magnetic Bragg peaks [(100) & (001)] and a broad peak is evident for both compounds over the temperature range 150-135 K for $x = 0.05$ and 140-125 K for $x = 0.1$. A similar coexistence of magnetic diffuse scattering and magnetic Bragg peaks below $T_N$ was reported for the layered $Sr_2CoO_3Cl$ compound and explained by a stacking disorder between antiferromagnetic sheets [30]. With decreasing temperature, the intensity of the broad peak decreases gradually and it disappears below ~135 K for $x = 0.05$ and ~125 K for $x = 0.1$. Interestingly, a small broadening of the magnetic Bragg peaks was also observed over these temperature ranges, as discussed earlier (figure 4). A broadening of a 3D magnetic Bragg peak occurs when correlations extend over a length scale smaller than the upper length scale given by the resolution limit of instrument. Therefore, the crossover of magnetic ordering from the short-range to 3D long-range (termed as 3D RLR) occurs over a wide temperature window (over 150-135 K for $x = 0.05$ and 140-125 K for $x = 0.1$ samples) in these bi-layered systems.

Figure 8 shows the proposed magnetic phase diagram for the $Ca_{2.5-x}La_xSr_{0.5}GaMn_2O_8$ system in the $x$-$T$ plane, based on the present study as well as on previous report [3], showing the effect of electron doping by substitution of $La^{3+}$ at the $Ca^{2+}$ site. From our neutron diffraction study, the observed magnetic ordered states for all compounds are AFM in nature. At lower temperatures ($T \leq$ 140, 135, and 125 K for $x = $ 0, 0.05, and 0.1, respectively), a true 3D LR AFM ordering was observed for all compounds. Over the temperature range 140-160 K for $x = 0$, 135-150 K for $x = 0.05$, and 125-142.5 K for $x = 0.1$, the magnetic ordering is 3D RLR AFM in nature. In the higher temperature range, different types of magnetic orderings have been observed for the samples with different electron doping. The $x = 0$ sample shows a 2D LR AFM and a 3D SR AFM orderings over 160-170 K and 170-190 K, respectively. For the La-substituted compounds, $x = 0.05$ and 0.1, a short-range AFM ordering has been observed over whole temperature range (150-195 K for $x = 0.05$ and 142.5-195 K for $x = 0.1$) above $T_N$. A pure 3D nature of the short-range magnetic ordering is more evident for the higher electron doped $x = 0.1$ compound.

Now, we discuss about the possible mechanism of the observed changes in the magnetic correlation with electron doping (La substitution) at $T > T_N$. As discussed in our previous report [17], double exchange FM interactions start to appear when doping the pure AFM system

Ca$_{2.5}$Sr$_{0.5}$GaMn$_2$O$_8$ with electrons due to the availability of hopping ($e_g$) electrons. A ferromagnetic character in the electron doped samples appears due to the formation of FM clusters inside the AFM matrix. The volume phase fraction of the FM clusters was found to be small (~ 2.7 % for $x = 0.1$ sample) [17]. Therefore, no observable ferromagnetic scattering was found in our present neutron diffraction study. However, the presence of FM clusters and/or the competition between coexisting AFM and FM interactions limits the extension of the AFM matrix. As a result, a decrease in the correlation length of the 2D AFM matrix occurs with strengthening the FM interaction, namely with increasing the La concentration. In this process, when the size of the 2D AFM matrix is small enough, the in-plane correlation lengths become comparable to the out-of-plane ones, and the system becomes effectively three dimensional in nature. For the undoped sample Ca$_{2.5}$Sr$_{0.5}$GaMn$_2$O$_8$ ($x = 0$) with no FM interactions, the AFM matrix is 2D in nature with a $\xi_{2D}$ ~ 550 Å at 160 K ($T/T_N$ ~ 1.03) [3]. On the other hand, for the electron doped samples, $x = 0.05$ and 0.1, the 2D correlation length $\xi_{2D}$ becomes comparable with the correlation length along the perpendicular $b$ direction namely with the thickness of the bilayer (~ 8 Å). As a result, the system effectively behaves like a three dimensional ($\xi_{3D}$ ~ 9.5 Å [figure 6]) system.

With decreasing temperature, a gradual changes of the magnetic ordering occur, extending over a wide temperature range [figures 7 and 8], and with different characteristics for the parent and substituted compounds: from a 3D SR to a 2D LR then to a 3D LR ordering for $x = 0$, and a SR (possible 3D) to a 3D LR ordering for $x = 0.05$ and 0.1. The above points suggest the presence of structural disorders, and/or defects in these samples. In fact, the presence of a local structural disorder in the parent compound was reported in literature [18]. A neutron diffraction study provides information about the averaged crystal structure. Therefore, to study possible defects or disorders, other local probe such as high resolution transmission electron microscopy and high resolution electron diffraction are required. The origin of local structure variations can also be studied by carrying out high-resolution neutron and/or synchrotron X-ray diffraction experiments through the estimation of ($hkl$)-dependent strain broadening due to lattice disorder.

## 4. Conclusions

The effect of electron doping, by substituting $La^{3+}$ at the $Ca^{2+}$ site, on the magnetic correlations as well as magnetic ordering dimensionality in the bilayered Brownmillerite compounds $Ca_{2.5-x}La_xSr_{0.5}GaMn_2O_8$ ($x$ = 0.05 and 0.1) has been investigated. A decrease of the 3D long-range AFM ordering temperature ($T_N$) from 160 K (for $x$ = 0) to 142.5 K (for $x$ = 0.1) is observed with substitution. A strong effect of electron doping has been found above $T_N$. Here, the doped compounds ($x$ = 0.05 and 0.1) show a short-range AFM ordering (with possible dimensionality of three) over temperature ranges 150-195 K for $x$ = 0.05, and 142.5-195 K for $x$ = 0.1, whereas, a 3D short-range AFM ordering over 170-190 K and a 2D long-range AFM ordering over 160-170 K were reported for the parent compound. The competition between coexisting AFM and FM interactions, introduced by the increase in the $Mn^{3+}$ / $Mn^{4+}$ ratio, likely reduces the extension of the average AFM ordering in a given $ab$ plane. This yields the observed change in the magnetic correlations above $T_N$. The reduced strength of the effective AFM superexchange interaction leads to the observed decrease in the ordering temperatures ($T_N$). The possibility of tuning of the dimensionality of magnetic ordering as well as the magnetic correlation lengths with electron doping in these layered systems should fuel up further interest for fundamental studies in the area of low dimensional magnetism as well as designing suitable spintronics materials.

# References

[1]   Jongh L J d 1990 *Magnetic Properties of Layered Transition Metal Oxide* (Netherlands Kluwer Academic Publishers)

[2]   Fruchart O and Thiaville A 2005 *C. R. Physique* **6** 921

[3]   Yusuf S M, Teresa J M D, Algarabel P A, Mukadam M D, Mirebeau I, Mignot J-M, Marquina C and Ibarra M R 2006 *Phys. Rev. B* **74** 184409

[4]   Majlis N, Selzer S and Strinati G C 1992 *Phys. Rev. B* **45** 7872

[5]   Majlis N, Selzer S and Strinati G C 1993 *Phys. Rev. B* **48** 957

[6]   Duijn J v, Attfield J P, Watanuki R, Suzuki K and Heenan R K 2003 *Phys. Rev. Lett.* **90** 087201

[7]   Kimura T and Tokura Y 2000 *Annu. Rev. Mater. Sci.* **30** 451

[8]   Lynn J W, Clinton T W, Li W-H, Erwin R W, Liu J Z, Vandervoort K and Shelton R N 1989 *Phys. Rev. Lett.* **63** 2606

[9]   Zhang H, Lynn J W and Morris D E 1992 *Phys. Rev. B* **45** 10022

[10]  Wu S Y, Hsieh W T, Li W-H, Lee K C, Lynn J W and Yang H D 1994 *J. Appl. Phys.* **75** 6598

[11]  Wu S Y, Li W-H, Lee K C, Lynn J W, Meen T H and D.Yang H 1996 *Phys. Rev. B* **54** 10019

[12]  Kamenev K V, Balakrisnan G, Paul D M K and Mcintyre G J 2002 *High Pressure Res.* **22** 135

[13]  Tsujii N, Suzuki O, Suzuki H, Kitazawa H and Kido G 2005 *Phys. Rev. B* **72** 104402

[14]  Yoshida Y, Yurue K, Mitoh M, Kawae T, Hosokoshi Y, Inoue K, Kinoshita M and Takeda K 2003 *Physica B* **329-333** 979

[15]  Hashi K, Tsujii N, Shimizu T, Goto A, Kitazawa H and Ohki S 2007 *J. Magn. Magn. Mater.* **310** 1242

[16]  Okada S and Matoba M 2009 *J. Phys.: Conf. Ser.* **150** 042154

[17]  Bera A K and Yusuf S M 2010 *J. Appl. Phys.* **107** 013911

[18]  Battle P D, Blundell S J, Brooks M L, Hervieu M, Kapusta C, Lancaster T, Nair S P, Oates C J, Pratt F L, Rosseinsky M J, Ruiz-Bustos R, Sikora M and Steer C A 2004 *J. Am. Chem. Soc.* **126** 12517

[19] Battle P D, Blundell S J, Santhosh P N, Rosseinsky M J and Steer C 2002 *J. Phys.: Condens. Matter* **14** 13569–13577

[20] Borjan Z, Popović Z S, Šljivančanin Ž V and Vukajlović F R 2008 *Phys. Rev. B* **77** 212402

[21] Allix M, Battle P D, Frampton P P C, Rosseinsky M J and Ruiz-Bustos R 2006 *Journal of Solid State Chemistry* **179** 775–792

[22] Gillie L J, Palmer H M, Wright A J, Hadermann J, Tendeloo G V and Greaves C 2004 *J. Phys. and Chem. Solids* **65** 87

[23] *http://www.ill.eu/sites/fullprof/*

[24] Rodriguez-Carvajal J April 2005 *FULLPROF, LLB CEA-CNRS.*

[25] Bertaut E F 1963 *Magnetism* (New York, Academic Press)

[26] Bertaut E F 1968 *Acta Crystallogr. Sect. A* **24** 217

[27] Borjan Z, Popović Z S, Šljivančanin Ž V and Vukajlović F R 2008 *Phys. Rev. B* **77** 212402

[28] Giot M, Chapon L C, Androulakis J, Green M A, Radaelli P G and Lappas A 2007 *Phys. Rev. Lett.* **99** 247211

[29] Carling S G, Visser D, Hautot D, Watts I D, Day P, Ensling J, Gütlich P, Long G J and F.Grandjean 2002 *Phys. Rev. B* **66** 104407

[30] Knee C S, Price D J, Lees M R and Weller M T 2003 *Phys. Rev. B* **68** 174407

[31] Warren B E 1941 *Phys. Rev.* **59** 693

Table 1. The Rietveld refined unit cell parameters, ordered moments, and the agreement factors for the compounds $Ca_{2.5-x}La_xSr_{0.5}GaMn_2O_8$ ($x = 0.05$ and $0.1$) at 4.5 and 275 K.

| Parameters | $x = 0.05$ | | $x = 0.1$ | |
|---|---|---|---|---|
| | 4.5K | 275K | 4.5K | 275K |
| $a$ (Å) | 5.4264(4) | 5.4380(4) | 5.4287(4) | 5.4342(5) |
| $b$ (Å) | 11.3733(9) | 11.3752(10) | 11.3620(9) | 11.3757(10) |
| $c$ (Å) | 5.2929(2) | 5.3060(2) | 5.2958(3) | 5.3107(2) |
| $\mu_{Mn}$ ($\mu_B$/Mn) | 3.02(1) | -- | 3.01(1) | -- |
| $\chi^2$ (%) | 13.6 | 12.9 | 13.1 | 7.66 |
| $R_p$ (%) | 4.38 | 4.11 | 4.02 | 3.05 |
| $R_{wp}$ (%) | 5.87 | 5.28 | 5.09 | 4.03 |
| $R_{Bragg}$ (%) | 4.11 | 6.23 | 6.26 | 4.47 |
| $R_{mag}$ (%) | 3.20 | -- | 4.51 | -- |

Table 2. Irreducible representations of the group of the propagation vector $G_k$.

| Symmetry element of $G_K$ | 1 | 2 (0, 0, 1/2) | (0, y, z) | (x, 0, z) |
|---|---|---|---|---|
| $\Gamma_1$ | 1 | 1 | 1 | 1 |
| $\Gamma_2$ | 1 | 1 | -1 | -1 |
| $\Gamma_3$ | 1 | -1 | 1 | -1 |
| $\Gamma_4$ | 1 | -1 | -1 | 1 |

Table 3. Basis Vectors of position 4c for representation $\Gamma_1$, $\Gamma_2$, $\Gamma_3$, and $\Gamma_4$.

| IR | | Basis vectors | | | |
|---|---|---|---|---|---|
| | | (x, y, z) | (-x, -y, z+1/2) | (-x, y, z+1/2) | (x, -y, z) |
| $\Gamma_1$ | $\Psi_1$ | (1 0 0) | (-1 0 0) | (1 0 0) | (-1 0 0) |
| | $\Psi_2$ | (0 1 0) | (0 -1 0) | (0 -1 0) | (0 1 0) |
| | $\Psi_3$ | (0 0 1) | (0 0 1) | (0 0 -1) | (0 0 -1) |
| $\Gamma_2$ | $\Psi_1$ | (1 0 0) | (-1 0 0) | (-1 0 0) | (1 0 0) |
| | $\Psi_2$ | (0 1 0) | (0 -1 0) | (0 1 0) | (0 -1 0) |
| | $\Psi_3$ | (0 0 1) | (0 0 1) | (0 0 1) | (0 0 1) |
| $\Gamma_3$ | $\Psi_1$ | (1 0 0) | (1 0 0) | (1 0 0) | (1 0 0) |
| | $\Psi_2$ | (0 1 0) | (0 1 0) | (0 -1 0) | (0 -1 0) |
| | $\Psi_3$ | (0 0 1) | (0 0 -1) | (0 0 -1) | (0 0 1) |
| $\Gamma_4$ | $\Psi_1$ | (1 0 0) | (1 0 0) | (-1 0 0) | (-1 0 0) |
| | $\Psi_2$ | (0 1 0) | (0 1 0) | (0 1 0) | (0 1 0) |
| | $\Psi_3$ | (0 0 1) | (0 0 -1) | (0 0 1) | (0 0 -1) |

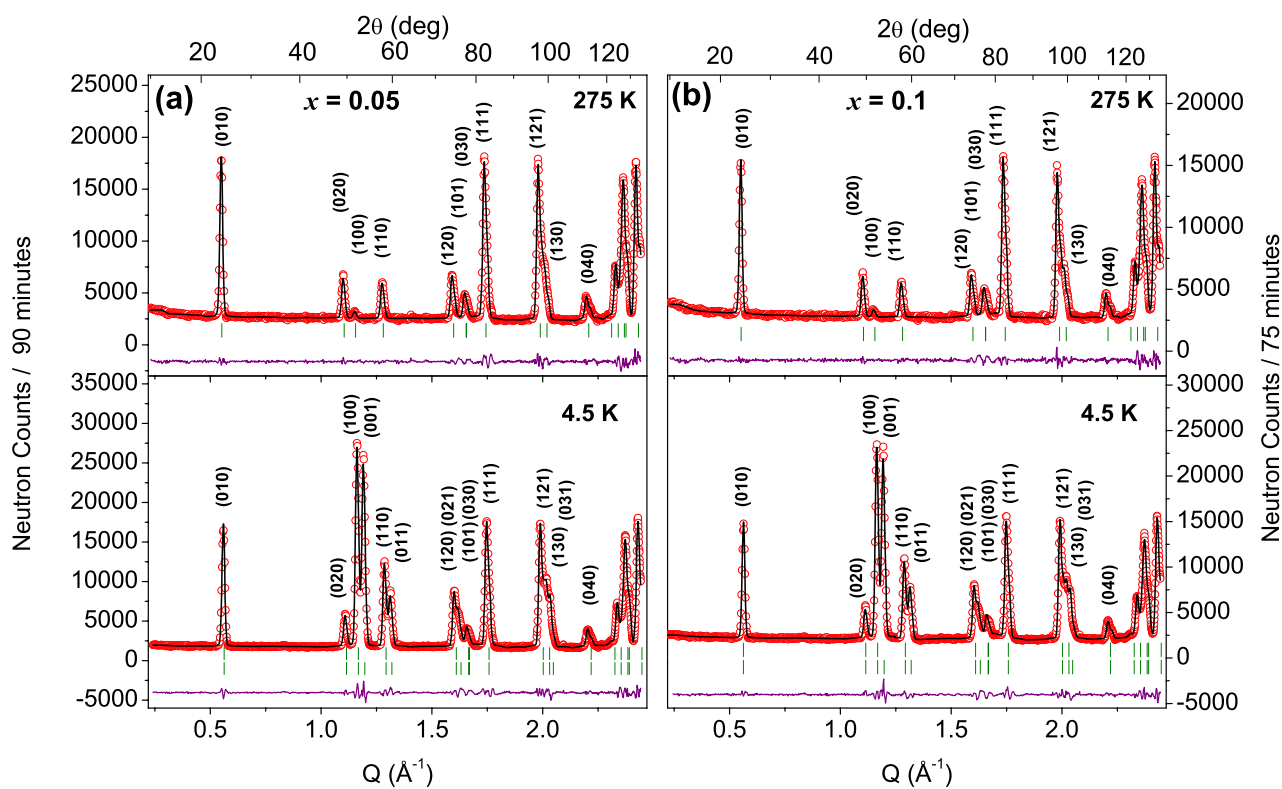

**Figure 1.** The observed (open circles) and calculated (solid lines) neutron diffraction patterns at 275 (paramagnetic state) and 4.5 K (AFM ordered state) for the compounds $Ca_{2.5-x}La_xSr_{0.5}GaMn_2O_8$ with (a) $x = 0.05$ and (b) 0.1. The solid lines at the bottom of each panel show the difference between observed and calculated patterns. The difference patterns are shifted downward for clarity. The vertical bars show the allowed nuclear (for both 275 and 4.5 K) and magnetic (for only 4.5 K, lower panel) Bragg peak positions. The (*hkl*) values of the observed nuclear (at 275 and 4.5 K) as well as magnetic (at 4.5 K) peaks are also listed.

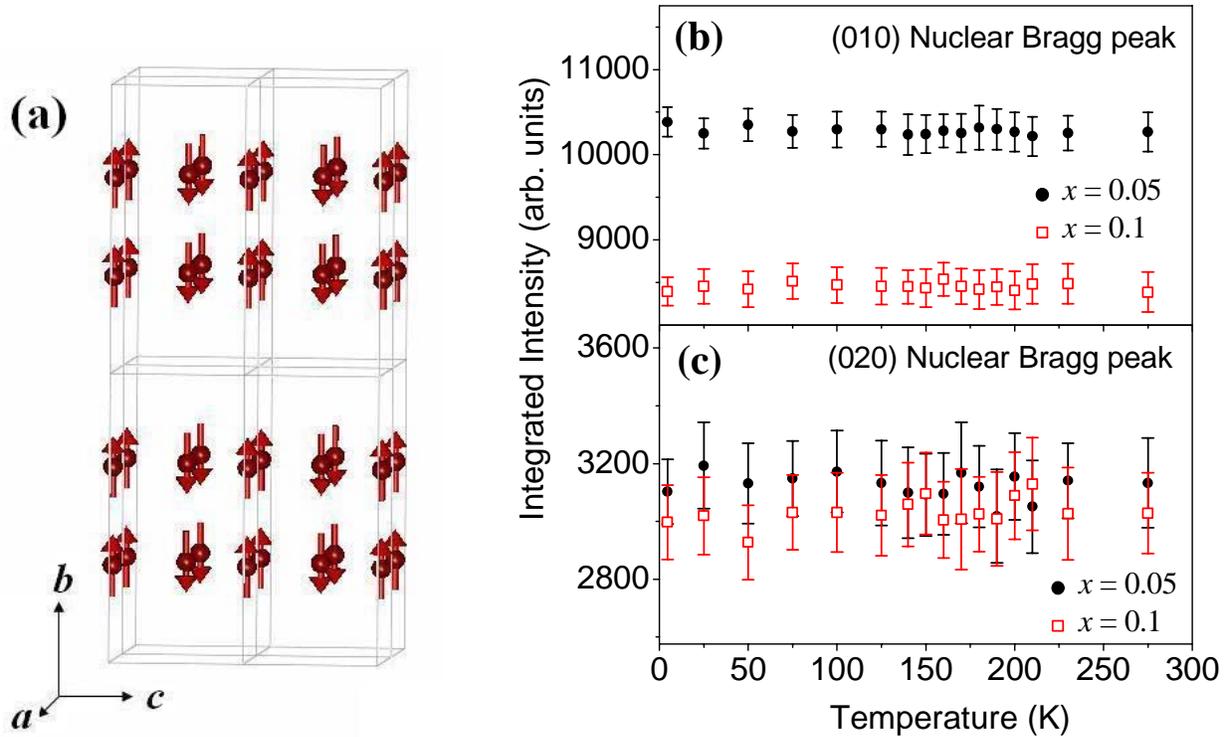

**Figure 2.** (a) The magnetic structure of the compounds $Ca_{2.5-x}La_xSr_{0.5}GaMn_2O_8$, corresponding to the representation $\Gamma_1$. For clarity, the magnetic structure is shown with a dimension of $2a \times 2b \times 2c$ where, $a$, $b$, and $c$ are nuclear unit cell parameters. (b) and (c): The temperature dependence of the integrated intensity of the (010) and (020) nuclear Bragg peaks, respectively, in $Ca_{2.5-x}La_xSr_{0.5}GaMn_2O_8$ ($x$ = 0.05 and 0.1). The data were collected for 38 and 65 mints. at each temperature for $x$ = 0.10 and 0.05 samples, respectively.

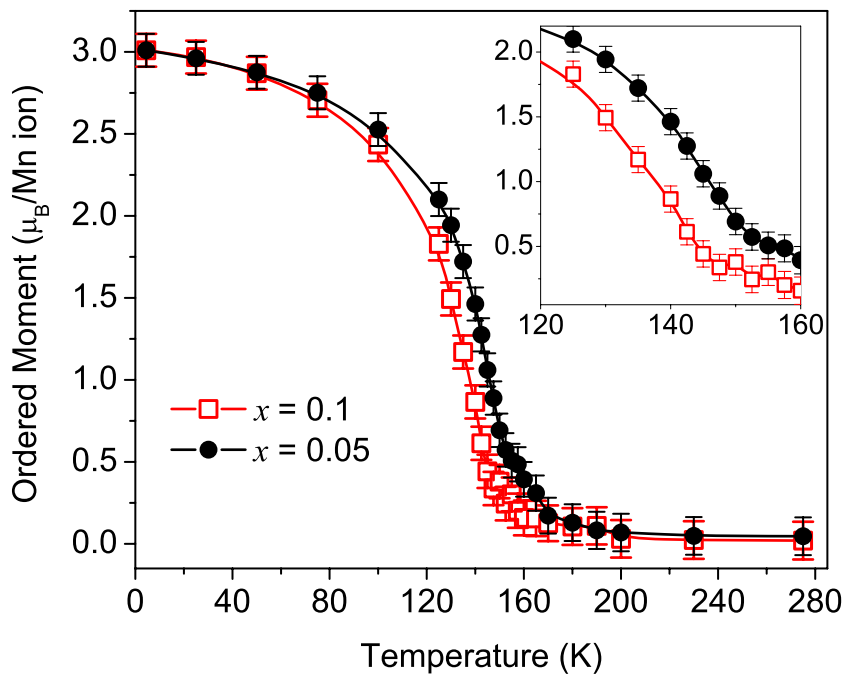

**Figure 3.** The temperature dependence of the refined magnetic moments for the compounds $Ca_{2.5-x}La_xSr_{0.5}GaMn_2O_8$ with $x = 0.05$ and $0.1$. The inset focuses on the transition region.

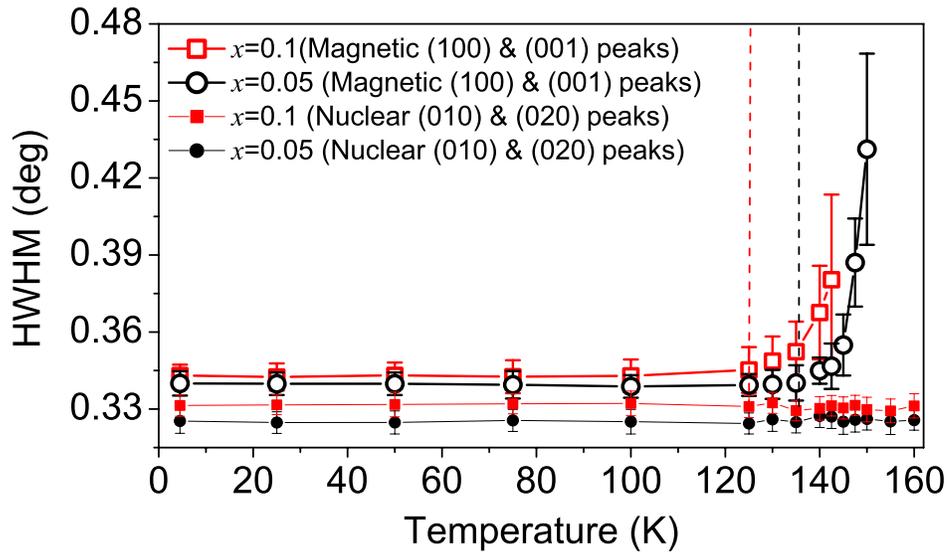

**Figure 4.** The temperature dependence of the average value of half width at half maxima (HWHM) of the 3D magnetic Bragg peaks (100) and (001) [open symbols] as well as pure nuclear Bragg peaks (010) and (020) [solid symbols] for the compounds $Ca_{2.5-x}La_xSr_{0.5}GaMn_2O_8$ with $x$ = 0.05 and 0.1.

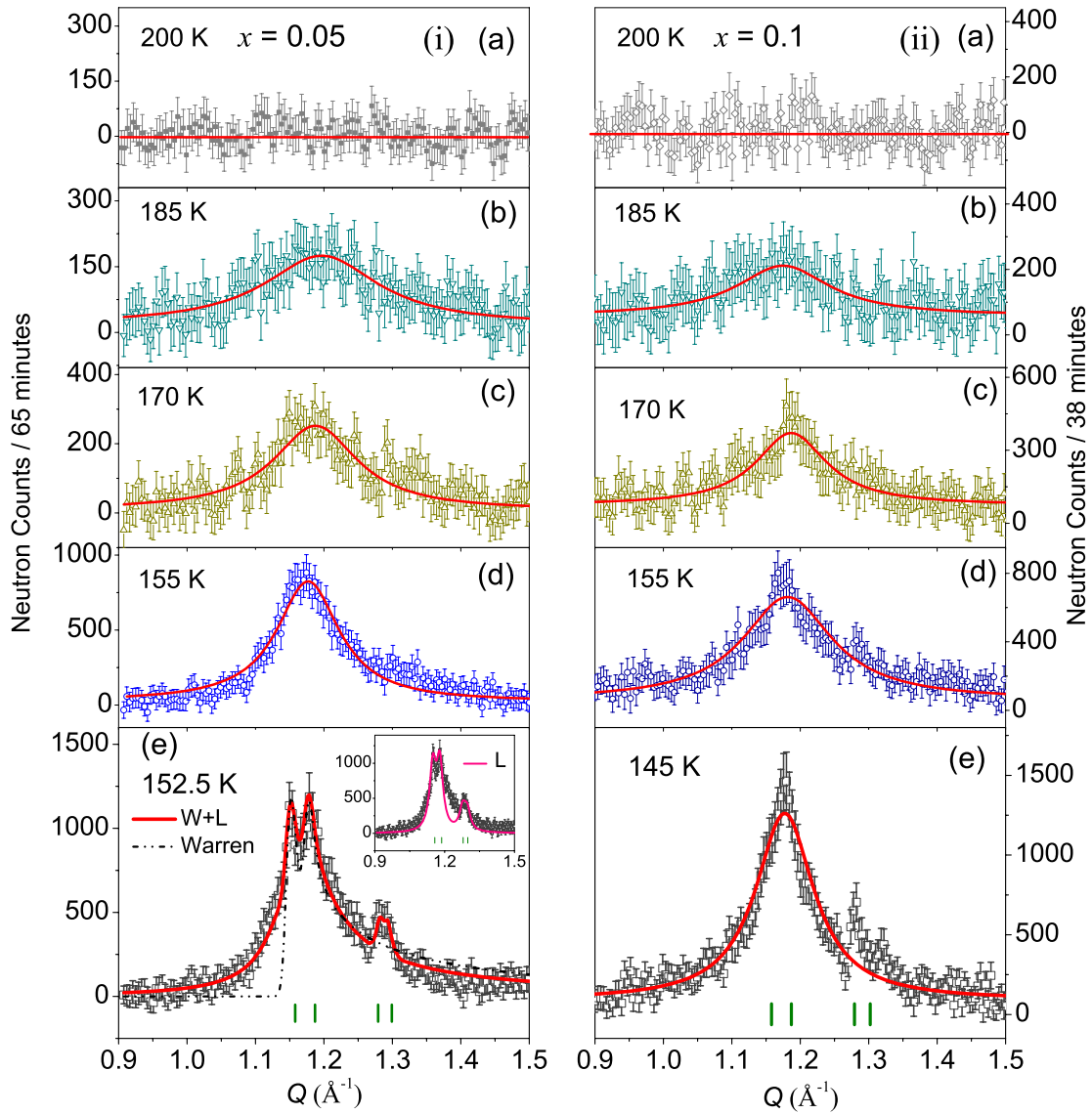

**Figure 5.** The magnetic diffraction patterns at (i) 152.5, 155, 170, 185, and 200 K for the compound with $x = 0.05$ and (ii) 145, 155, 170, 185, and 200 K for the compound with $x = 0.1$. The magnetic diffraction patterns are obtained after subtraction of the nuclear background at 210 K (paramagnetic state). Vertical bars show the positions of the (100), (001), (110), and (011) magnetic Bragg peaks below $T_N$. For the sample $x = 0.05$, the red thick curves in (i)(b)-(d) [185, 170, and 155 K] are the calculated profiles using a Lorentzian function. The flat lines (solid red) in (i)(a) and (ii) (a) indicate that no observable magnetic diffuse scattering is present at 200 K. For 152.5 K in (i)-(e), the red thick curve is the calculated profiles using a combined function of Warren and Lorentzian functions for each of the four magnetic peaks (at $Q \sim 1.15$ and $1.18$ Å$^{-1}$ and additional two peaks at $Q \sim 1.3$ Å$^{-1}$). The dotted curve is the calculated profile using Warren function alone for each of the two asymmetric Bragg peaks centred at $Q \sim 1.15$ and $1.18$ Å$^{-1}$. Inset in (i)-(e) shows the observed magnetic scattering at 152.5 K fitting with four Lorentzian functions centred at the position of the Bragg peaks as expected for a 3D behaviour. For the sample $x = 0.1$, red thick curves in (ii)(b)-(e) [185, 170, 155, 145 K] are calculated profiles using a single Lorentzian function.

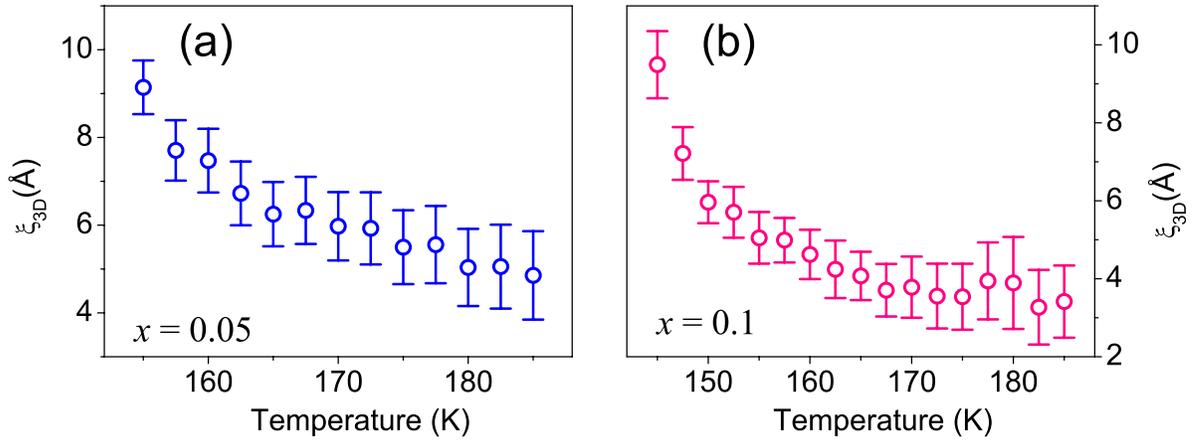

**Figure 6.** The temperature evolution of the 3D short-range spin-spin correlation length ($\xi_{3D}$) above $T_N$ for (a) $x = 0.05$ and (b) $x = 0.1$ samples.

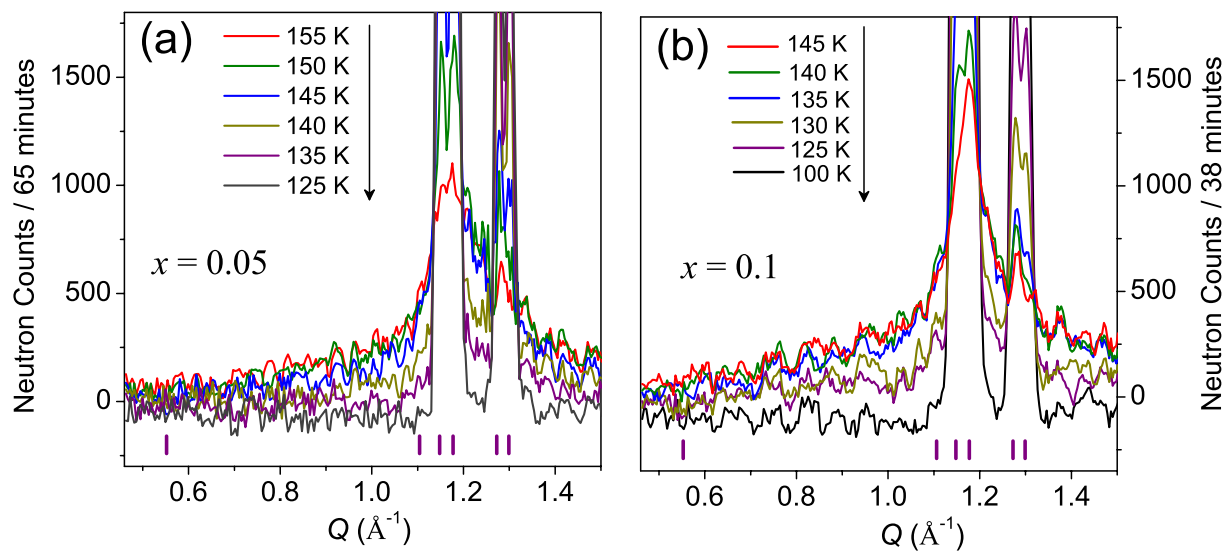

**Figure 7.** The magnetic diffraction patterns (after subtracting out the nuclear background at 210 K) at (a) 155, 150, 145, 140, 135 and 125 K for the sample $x = 0.05$ and (b) at 145, 140 135, 130, 125, and 100 K for the sample $x = 0.1$. The vertical bars show the positions of allowed nuclear as well as magnetic Bragg peaks.

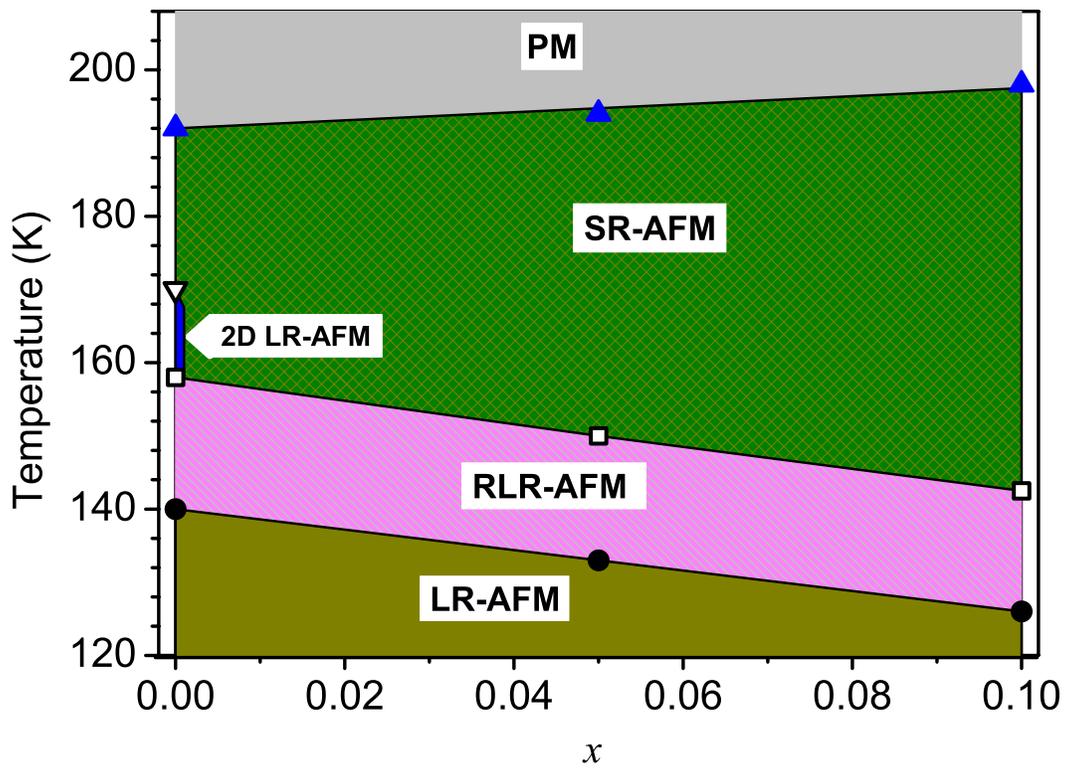

**Figure 8.** Magnetic phase diagram for the $Ca_{2.5-x}La_xSr_{0.5}GaMn_2O_8$ ($0 \leq x \geq 0.1$) systems in the $x$-$T$ plane: PM (Paramagnetic), SR (short-range), LR (long-range), RLR (restricted long-range). The observed ordering temperatures of the different magnetic states are shown for $x = 0$, 0.05, and 0.1 samples.